\documentclass[11pt]{article}

\usepackage{palatino}
\usepackage{mathpazo}
\usepackage{graphicx}
\usepackage{color}

\usepackage[margin=1in]{geometry}
\usepackage{amssymb,amsthm,amsmath}

\newtheorem{theorem}{Theorem}

\newtheorem{definition}{Definition}

\newcommand{\tr}{{\rm tr}}

\newlength{\blank}
\newenvironment{proof-of}[1][{\hspace{-\blank}}]{{\medskip\noindent\textit{Proof~{#1}.\ }}}{\hfill\qedsymbol}

\begin{document}

\title{\bf Additive entanglemement measures cannot be more than asymptotically continuous}
\author{
Andrea Coladangelo\footnote{%
     Computing and Mathematical Sciences,
     Caltech, Pasadena, California, USA.}
   \and
Debbie Leung\footnote{%
     Institute for Quantum Computing and
     Department of Combinatorics and Optimization,
     University of Waterloo,
     Waterloo, Ontario, Canada, and the Perimeter Institute.}
}

\date{22 October 2019}

\maketitle

\vspace*{-3ex}
\begin{abstract}
In this short note, we show that any non-constant quantity defined on
density matrices that is 
additive on tensor products and invariant under  
permutations cannot be ``more than asymptotically continuous''.  
The proof can be adapted to show that any additive entanglement
measure (on any number of parties) that is invariant under local
unitary operations also cannot be more than asymptotically 
continuous.
The proof is a direct consequence of generalizing a protocol in
\cite{LTW08} for embezzling entanglement.
\end{abstract}


We use ${\cal D}(d)$ to denote the set of density matrices on $d$
dimensions. These are $d \times d$ positive semidefinite matrices with
trace $1$. We use $\| \cdot \|_1$ to denote the Schatten $1$-norm of a
matrix; half of the Schatten $1$-norm of the difference between two
density matrices is the trace distance between the corresponding
quantum states. Let ${\cal D}(*) = \bigcup_{d} {\cal D}(d)$. We use
$\mathbb{R}_{+}$ to denote the set of non-negative real numbers.
\begin{definition}[Asymptotic continuity \cite{DHR01,SRH05}]
Let $f:  {\cal D}(*)  \rightarrow \mathbb{R}_{+}.$ 
We say that $f$ is asymptotically continuous if there exists a constant $K$ such that, for any $d$ and any 
$\mu, \nu \in {\cal D}(d)$,
\begin{equation}
|f(\mu) - f(\nu)| \leq K\, \| \mu - \nu \|_1 \log d + \eta(\| \mu - \nu \|_1)
\label{eq:asym-cty-def}
\end{equation}
where $\eta$ is independent of $d$ and vanishes with $\| \mu - \nu \|_1$.
\end{definition}
A well-known asymptotically continuous quantity is 
the Von Neumann entropy of density matrices 
$S(\rho) = -\tr \, \rho \log \rho$. This asymptotic continuity is, in turn, used to show
the asymptotic continuity of many other entanglement measures. 
Asymptotic continuity is often a key step in deriving expressions or
bounds for the asymptotic rates in quantum information processing
tasks, see \cite{MH97,THLD02} for examples. 

Examining the form of the continuity bound in \eqref{eq:asym-cty-def},
a ``more continuous'' quantity will vary more slowly with $\| \mu -
\nu \|_1$. 
\begin{definition}[``More than asymptotic'' continuity]
\label{def: more-than-ac}
Let $f: {\cal D}(*) \rightarrow \mathbb{R}_{+}.$ We say that $f$ is
more than asymptotically continuous if there exist $\alpha <1$ and $K$
such that, for any $d$ and any $\mu, \nu \in {\cal D}(d)$,
\begin{equation}
|f(\mu) - f(\nu)| \leq K\, \| \mu - \nu \|_1 (\log d)^{\alpha} + \eta(\| \mu - \nu \|_1)
\label{eq:more-asym-cty-def}
\end{equation}
where $\eta$ is independent of $d$ and vanishes with $\| \mu - \nu \|_1$.
\end{definition}

Throughout, we only consider non-constant quantities.  We say that $f$ is non-constant if there exist
$d\in \mathbb{N}, \rho, \sigma \in {\cal D}(d) ~\text{such that } |f(\rho) - f(\sigma)|  >0.$  
For composite systems, there are two additional natural properties.
First, for tensor product Hilbert spaces, corresponding to multiple
quantum systems (of not necessarily equal dimension), a quantity $f$ is
permutation-invariant if the quantity is independent of how the
systems are labeled.  
More precisely, for any $n \in \mathbb{N}$, any density operator
$\mu$ on $R = R_1 \otimes \cdots \otimes R_n$ and any 
permutation $\pi \in S_n$, it holds that 
$f(U_{\pi} \mu U_{\pi}^{\dagger}) = f(\mu)$,
where $U_\pi$ is the unitary transformation that permutes the $R_i$
registers according to $\pi$.
Second, a quantity $f$ is \textit{additive} (over tensor products)
if $f(\mu \otimes \nu) = f(\mu) + f(\nu)$ for all $\mu, \nu$. 
For example, the total spin number is both permutation-invariant 
and additive.  
These two natural properties are also satisfied by the entanglement entropy and by various other entanglement measures.  
%
%
We can now state our result about the relation between additivity, permutation-invariance and asymptotic
continuity.
\begin{theorem}
\label{onlyonethm}
Let $f:  {\cal D}(*)  \rightarrow \mathbb{R}_{+}$.
Suppose $f$ satisfies the following properties:
\begin{itemize}
\item additivity over tensor products, 
\item invariance under permutations, 
\item non-constant. 
\end{itemize}
Then, $f$ cannot be ``more than asymptotically continuous'', as  defined in 
Definition \ref{def: more-than-ac}.
\end{theorem}

Our proof is a simple consequence of an extension of a protocol from
\cite{LTW08} for embezzlement of entanglement, which can be described 
as follows.
Consider a scenario in which the direct transformation of a state
$\rho \in {\cal D}(d)$ into another state $\sigma \in {\cal D}(d)$ is
impossible. For example, the transformation could be forbidden because
$\rho$ and $\sigma$ contain different amounts of a quantity that is
conserved under the transformation.
For example, the $0$- and $1$-photon number states are not interconvertible 
if energy is conserved.  
Alternatively, the transformation can be impossible because the 
allowed set of operations could be limited.   
For example, if the transformations are limited to the set of local unitaries, 
and 
$\rho$ and $\sigma$ contain different
amounts of entanglement entropy, which is known to be conserved under local unitaries, then $\rho$ and $\sigma$ cannot be transformed into one another exactly.  
The transformation, although impossible to realize exactly, is known to be realizable approximately via a protocol from reference \cite{LTW08}, which uses a
``catalyst'' state $\Gamma \in {\cal D}(d^{n})$, where 
$n$ determines the accuracy of the transformation: 
\begin{equation}
\label{eq: gamma}
\Gamma = {1 \over n-1} \sum_{r=1}^{n-1} \rho^{\otimes r} \otimes
\sigma^{\otimes n-r}.
\end{equation}
Let $R$ be the system in which we want to transform $\rho$ to $\sigma$
and $R' = R_1 \otimes \cdot \cdot \cdot \otimes R_{n}$ be the system containing $\Gamma$,
with each $R_i$ being $d$-dimensional.  Let $\pi$ denote the cyclic 
permutation on $n$ elements $\pi: i \mapsto i+1 ~(\text{mod }n)$ and $U_{\pi}$ be the unitary transformation that cycles the registers according to $\pi$. Then,
$$U_{\pi} (\rho \otimes
\Gamma) (U_{\pi})^\dagger = \sigma \otimes \Gamma'$$ where 
\begin{equation}
\label{eq: gamma prime}
\Gamma' = {1 \over n-1} \sum_{r=1}^{n-1} \rho^{\otimes r+1} \otimes
\sigma^{\otimes n-r-1} = {1 \over n-1} \sum_{r=2}^{n} \rho^{\otimes r} \otimes
\sigma^{\otimes n-r}.
\end{equation}


Note that
$$ \Gamma - \Gamma' = {1 \over n-1} \left(\rho \otimes \sigma^{\otimes
  n-1} - \rho^{\otimes n} \right),$$ therefore,
$$ \| \Gamma - \Gamma' \|_1 \leq {2 \over n-1} \,.$$
So, the cyclic permutation $\pi$ achieves
the transformation 
$\rho \otimes \Gamma \rightarrow \sigma \otimes \Gamma'$
which approximates the forbidden, catalytic, transformation 
$\rho \otimes \Gamma \rightarrow \sigma \otimes \Gamma$ 
with accuracy
\begin{equation}
\| \sigma \otimes \Gamma' - \sigma \otimes \Gamma \|_1 =
\| \Gamma' - \Gamma \|_1 \leq {2 \over n-1}.
\label{eq:accbdd}
\end{equation}

We now return to the proof of Theorem \ref{onlyonethm}.  
Suppose for a contradiction that $f$ was more than asymptotically continuous. Pick $c,d>0$ and $\rho, \sigma \in {\cal D}(d)$ such that $|f(\rho) - f(\sigma)|= c$. Such $c, d, \rho , \sigma$ are guaranteed to exist since $f$ is non-constant by hypothesis. Consider 
$f(\rho \otimes \Gamma)$, 
$f(\sigma \otimes \Gamma')$, and 
$f(\sigma \otimes \Gamma)$ where $\Gamma$ and $\Gamma'$ are as defined 
in \eqref{eq: gamma} and \eqref{eq: gamma prime} respectively. We have
\begin{eqnarray}
c & = & |f(\rho \otimes \Gamma) - f(\sigma \otimes \Gamma)| 
= |f(\sigma \otimes \Gamma') - f(\sigma \otimes \Gamma)| 
\label{line1}
\\
& \leq & K\, \| \sigma \otimes \Gamma' - \sigma \otimes \Gamma \|_1 
(n+1)^\alpha (\log d)^\alpha + \eta(\|\sigma \otimes \Gamma' - \sigma \otimes \Gamma \|_1) 
\label{line2}
\end{eqnarray}
for some $\alpha<1$ and constant $K$, where the first equality of (\ref{line1}) comes from additivity of $f$, 
the second equality of (\ref{line1}) comes from permutation-invariance, 
and the inequality in (\ref{line2}) is by assumption.  Taking the large $n$ 
limit so that the $\eta$ term is negligible and applying (\ref{eq:accbdd}), 
we have
$$c \leq {2K \over n-1} (n+1)^\alpha (\log d)^\alpha$$ 
Since $\alpha<1$, the RHS tends to zero as $n$ goes to infinity, giving a contradiction. This concludes the proof of Theorem \ref{onlyonethm}.

Theorem \ref{onlyonethm} can be straightforwardly extended to entanglement measures. This is the content of Theorem \ref{entthm}. We introduce some notation. A partition $R$ of  $\mathbb{C}^d$ on $m$ parties is a decomposition of $\mathbb{C}^d$ into $\mathbb{C}^{d_1} \otimes \mathbb{C}^{d_2} \otimes
\cdots \otimes \mathbb{C}^{d_m}$ for some $d_1, d_2, \cdots, d_m$. We use
$R_i$ to label the $i$-th tensor component $\mathbb{C}^{d_i}$, and we write $R= R_1 \otimes R_2 \otimes \cdots \otimes R_m$. 
For two $m$-party partitions $R$ and $T$ (not necessarily with the same
dimension), where $R = R_1 \otimes R_2 \otimes \cdots \otimes R_m$, and
$T = T_1 \otimes  T_2 \otimes \cdots T_m$, the tensor product of
$R$ and $T$ is the partition $ R \otimes T = (R_1 \otimes T_1) \otimes (R_2 \otimes T_2) \otimes
\cdots \otimes (R_m \otimes T_m)$. 
For $n$ copies $R^{(1)}$, $\cdots$, $R^{(n)}$ of the same $m$-party partition $R$ (where we denote $R^{(i)} = R^{(i)}_1 \otimes R^{(i)}_2 \otimes \cdots \otimes R^{(i)}_m$), let the cyclic permutation
on the $j$-th party be the unitary transformation that cycles $R^{(1)}_j,
R^{(2)}_j, \cdots, R^{(n)}_j$.  The cyclic permutation on $R^{(1)} \otimes 
R^{(2)} \otimes \cdots \otimes R^{(n)}$ is the unitary $U_\pi$ formed by composing the cyclic
permutation for each party.

An entanglement measure on $m$-party states is a quantity defined on
the union of $m$-party partitions of $C^d$ for all $d$.  An entanglement measure
$f$ is additive (over tensor products) if for any partitions $R, T$, $\forall \rho \in R, \sigma \in T$, it holds that
$f(\rho \otimes \sigma) = f(\rho) + f(\sigma)$, where the LHS is
evaluated according to the partition of $R \otimes T$. An
entanglement measure $f$ is invariant under cyclic permutation if for any $\rho \in R^{(1)} \otimes 
R^{(2)} \otimes \cdots \otimes R^{(n)}$, where the $R^{(i)}$ are copies of some partition $R$,
$f(U_\pi \rho U_\pi^\dagger) = f(\rho)$. We now state a result similar to Theorem
\ref{onlyonethm} for entanglement measures.
\begin{theorem}
\label{entthm}
Let $f$ be an entanglement measure on $m$ parties for some natural number $m$.  
Suppose $f$ satisfies the following properties:
\begin{itemize}
\item additivity over tensor products, 
\item invariance under cyclic permutation, 
\item non-constant. 
\end{itemize}
Then, $f$ cannot be ``more than asymptotically continuous'', as  defined in 
Definition \ref{def: more-than-ac}.
\end{theorem}
The proof is essentially identical to that of Theorem \ref{onlyonethm} 
and we leave it to the interested readers.  
We remark that most entanglement measures are invariant under local
unitaries, which implies invariance under cyclic permutations.
%
%
The entanglement entropy in the bipartite setting is an example of
such an entanglement measure, but the result holds for any number of
parties, and any measure of mixed or pure state entanglement.

{\bf Acknowledgements}

AC is supported by the
Kortschak Scholars program and AFOSR YIP award number FA9550-16-1-0495.
DL is supported by NSERC.  

\newcommand{\etalchar}[1]{$^{#1}$}

\end{document}